\documentclass[useAMS,onecolumn]{mn2e}
\usepackage{epsfig}
\usepackage{amsmath,amssymb}
\usepackage{times}
\usepackage{color}
\usepackage{dcolumn}
\usepackage[english]{babel}

\begin{document}

\author[M. Shimon, Y. Rephaeli, N. Itzhaki, I. Dvorkin, and B.G. Keating]
{M. Shimon$^{1}$, Y. Rephaeli$^{1,2}$, N. Itzhaki$^{1}$, I. Dvorkin$^{1}$, 
and B.G. Keating$^{2}$\\
\\
$^1$ School of Physics and Astronomy, Tel Aviv University, 
Tel Aviv 69978, Israel\\
$^2$ Center for Astrophysics and Space Sciences, University
of California, San Diego, 9500 Gilman Drive, La Jolla, CA, 92093-0424\\}

\date{June 24, 2012}

\title[Constraints on the Neutrino Mass from SZ Surveys]
{Constraints on the Neutrino Mass from SZ Surveys}

\maketitle

\begin{abstract}

Statistical measures of galaxy clusters are sensitive to neutrino 
masses in the sub-eV range. We explore the possibility of using cluster 
number counts from the ongoing PLANCK/SZ and future cosmic-variance-limited 
surveys to constrain neutrino masses from CMB data alone. 
The precision with which the total neutrino mass can be determined from SZ 
number counts is limited mostly by uncertainties in the cluster 
mass function and intracluster gas evolution; 
these are explicitly accounted for in our analysis. 
We find that projected results from the PLANCK/SZ survey can be used to determine 
the total neutrino mass with a ($1\sigma$) uncertainty of $0.06$ eV, assuming  
it is in the range $0.1-0.3$ eV, and the survey detection limit is set at the 
$5\sigma$ significance level. Our results constitute a significant improvement on 
the limits expected from PLANCK/CMB lensing measurements, 0.15 eV. 
Based on expected results from future cosmic-variance-limited (CVL) 
SZ survey we predict a $1\sigma$ uncertainty of $0.04$ eV, a level comparable 
to that expected when CMB lensing extraction is carried out with the same experiment. 
A few percent uncertainty in the mass function parameters could result in up 
to a factor $\sim 2-3$ degradation of our PLANCK and CVL forecasts.
Our analysis shows that cluster number counts provide a viable complementary 
cosmological probe to CMB lensing constraints on the total neutrino mass.

\end{abstract}

\maketitle

\section{Introduction}
CMB measurements already placed meaningful upper limits on the total neutrino mass 
from its impact on the early integrated Sachs Wolfe (ISW) effect. The energy scale 
of recombination, $\sim 0.3$ eV, sets the value of this upper limit; if the total 
neutrino mass is larger than this value, then neutrinos are non-relativistic 
and do not contribute to the decay of gravitational potentials shortly after 
recombination. If, on the other hand, the total mass is lower they constitute 
a relativistic component that contributes to the 
decay of linear gravitational potentials, which causes a net change in the 
temperature of the CMB towards these gravitational wells.

Measurements of CMB polarization open yet another window on neutrino masses via CMB 
lensing by the intervening large scale structure at redshifts of a few. 
Multiple ongoing ground-based CMB experiments are targeting this lensing-induced 
B-mode signal which is expected to peak on a few arcminute scales , $l\sim 1000$, 
far away from the predicted, much weaker primordial (inflation-induced) B-mode 
signal that is expected to peak at degree scales, $l\sim 100$.
It has been shown that applying optimal estimators to CMB temperature and polarization 
maps one can recover the lensing potential to the precision that will allow constraining 
the total neutrino mass to the $0.04$ eV level (Kaplinghat, Knox \& Song, 2003) 
with a cosmic-variance-limited (CVL) CMB experiment, assuming full-sky coverage, 
no foregrounds, and no source of non-gaussianity other than the lensing of the CMB, 
a feature on which the optimal filters rely. In practice, it is unlikely that all 
these conditions will be fully satisfied and in that sense the frequently-quoted 
value $0.04$ eV is quite likely too optimistic. Yet, this is currently the most 
promising CMB-based probe for neutrino mass inference.

Other cosmological neutrino probes can constrain their masses 
to varying degrees, e.g. Abazajian et al. (2011), Wong (2011), 
and supplementary laboratory experiments which are expected to
reach the $\sim$0.2 eV sensitivity. Complementary cosmological probes
of neutrino masses include: CMB temperature anisotropy (e.g., MacTavish et al. 2006, 
Komatsu et al. 2010), weak lensing and shear maps (e.g., Cooray 1999, 
Abazajian \& Dodelson 2003, Song \& Knox2004, Hannestad, Tu \& Wong 2006, 
Kitching et al. 2008, Namikawa et al. 2010), galaxy 
(Hu, Eisenstein \& Tegmark 1998, Hannestad 2003, Tegmark et al. 2004,
Tegmark et al. 2006, Thomas, Abdalla \& Lahav 2010  ) and Ly$\alpha$ surveys 
(Croft, Hu \& Dave 1999, Goobar et al. 2006, Seljak, Slosar \& McDonnald 2006, 
Gratton, Lewis \& Efstathiou 2008). Joint of these different datasets 
will meaningfully constrain the neutrino masses Joudaki \& Kaplinghat (2011). 
In order to asses the relative importance of the yields of these various probes, 
the upper limits that they set should be compared to the lower limits obtained 
on neutrino masses from neutrino oscillation experiments. The observed level 
of neutrino oscillations implies that at least one of the neutrinos is 0.05 eV 
or heavier. This picture corresponds to the 'normal hierarchy'. In the 'inverted 
hierarchy' two neutrino masses are each above 0.05 eV. This implies that the 
lowest bound on the total neutrino mass lies in the range 0.05-0.10 eV, which 
sets the benchmark level for determining the hierarchy, ultimately allowing 
rejection of one of these neutrino mass models. The goal is therefore to push 
the (cosmological) upper neutrino mass limits down below 0.05 eV. 

Cluster number counts are yet another useful probe of neutrino masses. This 
is due to the fact that typical cluster scales are much smaller than the 
$\sim 150$ Mpc scale of linear dark matter halos that lens 
the CMB. In addition, cluster number counts are exponentially sensitive to 
$\sigma(M,z)$, the rms mass fluctuation on the cluster mass scale $M$ at 
redshift $z$, and since $\sigma(M,z)$ itself is exponentially sensitive to 
neutrino mass (via the growth function, see e.g. Eq.134 of Lesgourgues \& Pastor 2006) 
this implies that cluster number counts should be a rather sensitive probe of neutrino 
masses, as has already been demonstrated, e.g., Wang et al. (2005), 
Shimon, Sadeh \& Rephaeli (2011); hereafter, SSR, and recently also 
Carbone et al. (2011). While the latter work is also based on cluster number counts,
our treatment here includes explicit accounting for uncertainties in
the (all important) cluster mass function, and a wider range of fiducial neutrino 
masses (0.1-0.3 eV). Here we further explore the ability 
to strengthen the constraints on the neutrino mass from cluster number counts. 
We do so by parameterizing uncertainties in 
the halo mass function, which is the dominant source of modeling uncertainties. 
This important function, whose specific shape and normalization reflect the 
details of the growth of density fluctuations, and the non-linear 
hierarchical collapse and mergers of sub-structures, can be best studied 
by state-of-the-art large-volume hydrodynamical cosmological simulations. 
Although very advanced, currently available numerical 
codes predict a range of mass functions. At present, this mass function 
indeterminacy largely sets the precision limit of forecasting the total 
neutrino mass from cluster SZ number counts and power spectrum. In the first 
phase of our work (SSR) we considered highly degenerate nuisance parameter which 
greatly degraded our results. Aside from 
reconsidering this issue, in the current work we also include cluster sample variance 
errors (in addition to Poissonian noise), a more realistic intracluster (IC) gas 
profile, as well as gas evolution with cluster mass and redshift.

Our basic approach and details of the Fisher Matrix analysis are only briefly 
described here; a more extensive description can be found in SSR.  
In section 2 we discuss the impact of massive neutrinos on the LSS and 
cluster number counts, and briefly describe, in section 3, the Fisher matrix 
analysis used in this work. Our main results are presented in section 4, and 
further discussed in section 5.

\section{Neutrino Impact on Growth of the LSS}

The evolution of structure in the matter-dominated era is embodied in 
the matter power spectrum,
\begin{eqnarray}
P_{m}(k,z)=Ak^{n}T^{2}(k,z),
\end{eqnarray}
where $Ak^{n}$ is the primordial density fluctuation spectrum, with $A$ its 
overall normalization, $n$ is the tilt of the power spectrum, and 
$T(k,z)=T(M_{\nu};k,z)$ is the transfer function. An important quantity gauging 
the amplitude of the processed power spectrum observed today is the mass variance 
parameter on a scale of ${\rm 8 \,Mpc\ h^{-1}}$, 
\begin{eqnarray}
\sigma_{8}^{2}=\int_{0}^{\infty}P_{m}(k,z)W^2(kR)k^2\frac{dk}{2\pi^{2}},
\end{eqnarray}
where $W(kR)$ is a window function, and $R ={\rm  8 \,Mpc\ h^{-1}}$. 
The latter quantity incorporates the physics of neutrino damping; thus, $\sigma_{8}$ 
is a function of not only $A$ and $n$, but also of neutrino masses (via the 
transfer function), and indeed any other cosmological parameter which 
affects structure evolution on scales of few tens of Mpc and below. Because 
these scales are comparable to the typical scales of diffusion damping of 
density fluctuations by neutrinos with small masses, we expect that 
$M_{\nu}$ and $\sigma_{8}$ will be anti-correlated. Since the SZ signature 
is a strong function of $\sigma_{8}$, it is expected to be sensitive also 
to $M_{\nu}$. More generally, the mass variance on a physical scale $R$ and at 
redshift $z$ is
\begin{eqnarray}
\sigma_{R}^{2}(z)=\int_{0}^{\infty}P_{m}(k,z)W^2(kR)k^2\frac{dk}{2\pi^{2}}
\end{eqnarray} 
where the  effective upper cutoff of the integral 
is $\sim R^{-1}$. For large $R$ there generally is little impact 
of $M_{\nu}$ on $\sigma_{R}$, but on cluster scales, where $R$ is 1 Mpc or smaller, 
the impact of non-vanishing $M_{\nu}$ on $\sigma_{R}$ is considerable. 
As we show below, the discriminative power of cluster number counts stems 
from the exponential dependence of the mass function on $\sigma_{R}$.
In our numerical calculations of the SZ effect we employ a publically 
available code (Kiakotou, Elgaroy \& Lahav 2007) to determine 
$T(k,z)$ assuming three degenerate neutrino masses, but we do use 
the default CAMB transfer function to calculate the primordial 
angular power spectrum of the CMB. This is warranted since 
the main difference between the two transfer functions is most apparent at 
large values of $k$. 

The basic quantity which describes the 
number density of clusters as a function of their mass and 
redshift - basic properties by which clusters are identified - 
is the mass function. As will be shown later, the total neutrino mass can be 
derived from comparison of the observed number of clusters in a given redshift-bin 
to the number predicted from the mass function, $\frac{dn(M;z)}{dM}$. 
The latter function is defined in terms of the differential number of clusters in 
a volume element $dV$,
\begin{eqnarray}
dN(M,z)=f_{sky}\frac{dn(M,z)}{dM}dVdM ,
\end{eqnarray}
where $f_{sky}$ is the observed sky fraction, which for the two experiments discussed 
here is 0.65 when realistic masking of the galaxy is considered. 
The total number in a given interval $\Delta z$ around $z_{i}$ is
\begin{eqnarray}
\Delta N(z_{i})=f_{sky}\Delta z_{i}\frac{dV(z_{i})}{dz}\int\frac{dn(M,z_{i})}{dM}dM.
\end{eqnarray}
As noted earlier, the current most optimal determination of the mass function is 
from cosmological simulations. 
Here we adopt the mass function derived recently by Tinker et al. (2008) 
from a large set of dynamical cosmological simulations in the $\Lambda$CDM scenario. 
Expressing the mass function in the conventional form, 
\begin{eqnarray}
\frac{dn}{dM}=f(\sigma)\frac{\rho_{m}}{M}\frac{d\ln(\sigma^{-1})}{dM} ,
\end{eqnarray}
these authors derived a fit of the form 
\begin{eqnarray}
f(\sigma)=A\left[\left(1+\frac{\sigma}{b}\right)^{-a}\right]
e^{-\frac{c}{\sigma^{2}}} ,
\end{eqnarray}
where the parameters $A$, $a$, $b$ and $c$ were the best-fits to the results of their 
simulations. The mass function does not have a universal form, a fact 
that becomes apparent by the deduced dependence of these fit parameters on 
both redshift and the overdensity at virialization, $\Delta_{v}$
\begin{eqnarray}
A&=&A_{0}(1+z)^{-0.14}\nonumber\\
a&=&a_{0}(1+z)^{-0.06}\nonumber\\
b&=&b_{0}(1+z)^{-\alpha}\nonumber\\
\log(\alpha)&=&-\left(\frac{0.75}{\log(\Delta_{v}/75)}\right)^{1.2}\nonumber\\
b_{0}&=&1.0+(\log(\Delta_{v})-1.6)^{-1.5}
\end{eqnarray}
where $c$ and $A_{0}$ are obtained from Table 2 of Tinker et al. (2008) 
which we also used in deriving fits for $a_{0}$ and $b_{0}$ 
as functions of $\Delta_{v}$. The following are very good fits for the relevant 
$\Delta_{v}<400$ range
\begin{eqnarray}
a_{0}&=&1.7678\alpha_{1} -0.5941\alpha_{2}\exp(-0.02924\Delta_{v}^{0.5967})\nonumber\\
c&=&1.7077\alpha_{3} -0.7038\alpha_{4}\exp(-0.001\Delta_{v}^{1.079})
\end{eqnarray}
where $\alpha_{1}-\alpha_{4}$ equal $1$ in the fiducial (Tinker et al. 2008) model.  
In our analysis we include these four nuisance parameters to account 
for possible small uncertainties or biases in these parameters. 

The SZ effect is a unique probe of cosmological parameters and cluster 
properties; its statistical diagnostic value is reflected through number 
counts and the power spectrum of the CMB anisotropy it induces. As we have stated 
already, the dependence of cluster number counts on the total neutrino mass, and an 
assessment of the feasibility of actually determining it from cluster surveys 
are our main objectives in this work. The details of the calculation of the statistical 
SZ signal from individual z-shells, i.e. clusters residing within a 
narrow redshift interval, follow those described in SSR.  
We adopt the spherical collapse model for cluster formation,
and we relate the virial cluster mass to its radius 
using the standard relation.

DM profiles are assumed to have the 
the Navarro-Frenk-White (NFW; Navarro, Frenk \& White 1995) form, 
with the mass-concentration relation $c(M,z)$ from Duffy et al. (2008). 
We assume a polytropic equation of state for 
IC gas with a polytropic index $\Gamma=1.2$. The solution of the equation
of hydrostatic equilibrium for a polytropic gas inside the potential well of a
DM halo is (Ostriker, Bode \& Babul 2005):
\begin{equation}
 \rho(x)=\rho_0\left[1-\frac{B(\Gamma-1)}{\Gamma}\left(1-
\frac{\ln(1+x)}{x} \right)\right]^{1/(\Gamma-1)}
\label{eq:rhox}
\end{equation}
where $x=r/r_s$, $r_s$ is the scale factor of the NFW density profile, $B$ is
given by $B=4\pi G\rho_s r_s^2\mu m_p/k_B T_0$ and $\mu m_p$ is the mean
molecular weight. Further details of the IC gas model used in this work are
given in Dvorkin, Rephaeli \& Shimon (2012).

IC gass mass fraction is assumed to follow 
that of Vikhlinin et al. (2009) with a nuisance parameter in our analysis
\begin{eqnarray}
f_{g}(M,z)=\alpha_{5}[0.125+0.037\log_{10}(M_{500}/M_{15})]
\end{eqnarray}
with fiducial value $\alpha_{5}=1$.
Here $M_{500}/M_{15}$ is the total cluster mass enclosed in a 
sphere of overdensity of 500 in units of $10^{15}$ solar masses. The 
redshift dependence in Eq.(11) is that of $M_{*}$, defined such that 
for a fixed redshift the mass fluctuation $\sigma(M_{*},z)=1.686$. 
Under these assumptions we normalize the SZ power spectrum to conform 
with the value measured recently by SPT (Reichardt et al. 2011), 
$C_{l}=3.65 \mu K^{2}$ at $l=3000$. We do so for each fiducial mass separately 
such that for each of the three fiducial values considered here 
(0.1, 0.2, and 0.3 eV) the power spectrum at $l=3000$ is the same.

Nonetheless, the shape of the SZ power spectrum 
depends on the fiducial neutrino mass, but the power level at the relevant range 
$2000<l<3000$ is nearly the same, implying that a similar number of galaxy clusters 
is expected to be detected (for a fixed SPT level at $l=3000$), resulting in neutrino 
mass uncertainties essentially independent on the assumed fiducial neutrino mass.

\section{Fisher Matrix Analysis}

Our cosmological model includes the normalization $A$ and tilt $n$
of the primordial scalar perturbations, neutrino mass $M_{\nu}$, 
matter, $\Omega_{m}$, and baryon, $\Omega_{b}$, density parameters, 
the Hubble parameter (scaled to $100$ km/sec/Mpc) $h$, 
dark energy equation of state $w$, optical depth to reionization $\tau$, and the 
primordial helium abundance $Y_{p}$. The priors on the cosmological parameters are 
obtained from the primary and lensed sky observed with PLANCK and the CVL experiments, 
with the corresponding Fisher matrices denoted by $F^{pr}$ and $F^{LE}$, 
respectively. We also use the prior $H_{0}=71.0\pm 2.5$ 
km/sec/Mpc. 
In calculating the Fisher matrix for the primary CMB (with and without 
lensing extraction, LE) we follow the standard approach which we do not reproduce here; 
details of the calculation can be found in, e.g., Lesgourgues et al. (2006).
In addition to the above nine cosmological parameters we included two free parameters in 
the gas mass fraction Eq.(10) and four parameters to describe 
small departures from the values of Tinker et al. (2008; see our eqs. 8-9).

The likelihood function for cluster number counts would naively be written as a Poisson 
distribution involving the observed and theoretical number counts in 
redshift-shells (e.g., Holder, Haiman \& Mohr 2001). 
In practice, however, for higher abundance clusters one has to include also 
the cluster correlation term that increases the naive sample bias (Hu \& Kravtsov 2003). 
Our $5\sigma$ detection threshold guarantees that only the most massive clusters will 
be detected, rendering the effect of this extra term small; our estimations of 
neutrino mass uncertainty from the combined primordial CMB and number counts 
increase by only $10-15\%$ when including this extra term.

We set the shell-width to $\Delta z=0.1$ which is safely larger than predicted 
photo-z redshift uncertainties which are at the $\sigma_{z}=0.02(1+z)$ level. 
We found that increased refinement of the redshift bins essentially 
does not affect our results. This improvement never exceeds the $15\%$ level; 
therefore, we adopted $\Delta z=0.1$. The cutoff on our cluster sample was set at 
$z_{max}=1.0$ because we found that removing individual redshift shells (with a fixed 
CMB prior on the cosmological parameters) does not degrade our results 
beyond $z\approx 0.2-0.3$; most of the constraint on masses comes from low-redshift 
high-mass clusters. We also explored the possibility of using the CVL and PLANCK 
catalogs of thousands clusters to analyze the data in $M-z$ cells and found that 
doing so (if mass bins are sufficiently wide to allow a factor $\sim 3$ uncertainty 
in cluster mass determination) will add very little to the constraining power of 
our analysis due to our requirement that each cell contains at least 20 clusters 
and the fact that clusters in our sample are very much uniform in size and redshift, 
and hence also in mass. In addition, there is a strong correlation 
between the mass and redshift of detected clusters. This correlation is most apparent 
for high-z clusters; when such a cluster is detected, it mostly lies within a narrow 
mass range just above the threshold for detection. At lower redshifts the mass 
range of detected clusters increases but the strong M-z correlation persists. Therefore, 
counting galaxy clusters in redshift-bins essentially amounts to counting clusters 
in mass bins which is indeed a more direct probe of neutrino mass. The only 
difference being that cluster mass inference is highly model-dependent while their 
redshift determination is much cleaner.

The approximate Fisher matrix for cluster number counts reads (Lima \& Hu 2004)  
\begin{eqnarray}
F_{\mu\nu}^{N}=\sum_{i,j}N_{,\mu}^{i}(C^{-1})_{ij}N_{,\nu}^{j}
+\frac{1}{2}{\rm Trace}[C^{-1}S_{,\mu}C^{-1}S_{,\nu}]
\end{eqnarray}
where $C=N+S$ is the total covariance matrix; $N$ is the diagonal Poissonian 
part while $S$ encodes the correlation between redshift bins and generally has 
small off-diagonal terms, but is less relevant for our case of wide bins, 
and more generally when correlations between populations in different bins 
are negligible. The estimated uncertainty in the parameter $\lambda_{\mu}$ is then
\begin{eqnarray}
\Delta\lambda_{\mu}=(F_{\mu\mu}^{N})^{-1/2} ,
\end{eqnarray}
where we take the square root of the $j$'th Fisher matrix element. 
The total Fisher matrix that includes cluster number counts combined 
with either the primordial or lensed CMB is $F_{\mu\nu}^{pr}+F_{\mu\nu}^{N}$ or 
$F_{\mu\nu}^{{\rm LE}}+F_{\mu\nu}^{N}$, respectively, and the uncertainty in the 
parameter $\lambda_{\mu}$ is similar to that in Eq.(13).
To estimate the signal-to-noise $\mathcal{S/N}$ with which a cluster can 
be detected in a survey we follow our original work and assume that main sources of 
noise are instrumental, primary CMB anisotropy, and point source contamination. We assume 
the performance of optimal matched filters as applied in SSR in order 
to estimate the abundance of detected clusters on a finely-sampled $M-z$ grid 
(sampled to the level required for convergence of our constraints on neutrino mass).

\section{Results}

In this work we have adopted the $\Lambda$CDM cosmological model with 
WMAP-7 best-fit parameters. The cluster population is described in terms of 
the mass function of Tinker et al. (2008), with IC gas mass fraction 
described by Eq.(11) and IC gas profile as in Eq.(\ref{eq:rhox}). 
It is the high sensitivity of the halo mass function to neutrino masses 
(e.g. Brandbyge et al. 2010) that we use as a probe of the total neutrino mass.
It is important to note that in this analysis no priors were set on 
the cosmological parameters, nor on the gas mass fraction 
(modeled in our analysis with one parameter), 
except for $H_{0}$ for which we 
adopted an uncertainty of 2.5 km/sec/Mpc. In addition, we considered four 
nuisance parameters that describe possible departures from the Tinker mass 
function and explored the robustness of $\sigma_{M_{\nu}}$ to small changes 
in these parameters ($\alpha_{1}-\alpha_{4}$) assuming that these parameters 
are known at the 1-10\% precision level. 

PLANCK specifications are given in Table 1. For $F^{pr}$ and $F^{LE}$ we 
used all nine frequency bands while for simulating cluster detection 
for our cluster number counts analysis we assumed only the 100, 143 and 353 
GHz bands are used.
Our basic results for the neutrino mass uncertainty based on cluster number 
counts from SZ surveys with the ongoing PLANCK and CVL 
experiments are presented in Table 2. 
We show (from left to right) the expected uncertainty on the inferred $M_{\nu}$ 
from the primary CMB (both temperature anisotropy and polarization), lensing 
extraction (LE) of the CMB, primary CMB and cluster SZ number counts, 
LE and cluster number counts, and finally the total number of clusters expected 
to be detected. The cluster mass range we considered is 
$3\times 10^{13}M_{\odot}-3\times 10^{15}M_{\odot}$, but we verified that 
the high $\mathcal{S/N}$ detection threshold that we imposed on the cluster 
samples guarantees that the detected detected clusters are much more massive than 
the imposed lower mass bound.

The results presented in Table 2 demonstrate that with cluster number counts 
alone (and priors based on measurements of the primary CMB power spectrum 
and the HST prior on $H_{0}$) neutrino mass uncertainties may be constrained 
to the $\sim 0.04-0.06$ eV range, depending on the details of the SZ cluster 
surveys. The $\sim 0.04-0.06$ eV range brackets 
our expectations assuming we trust the mass function - a standard practice in 
this approach. To test the robustness of our estimates to possible deviations 
from values of the parameters in the analytic representation of the mass function 
adopted here, we allowed some freedom in these parameters, beyond the fiducial 
values that were obtained from best-fitting to simulation (Tinker et al. 2008). 
A few examples that illustrate our findings are as follows: 
We find that the primordial CMB combined with PLANCK 
cluster number counts give $\sigma_{M_{\nu}}=0.06$ eV (using the Tinker 
mass function). With a 10\% uncertainty in the four mass function nuisance 
parameters, the uncertainty grows to $\sigma_{M_{\nu}}=0.12$. 
Similarly, the CVL SZ survey will result in $\sigma_{M_{\nu}}=0.04$ or $0.11$ eV 
if the mass function is precisely known or known to only 10\% precision, respectively. 
The conclusion from this is that accounting for mass function uncertainty is particularly 
important when reliable constraints on neutrino mass are required.

\section{Discussion}

The CMB is perhaps one of the best understood and most model-independent 
underpinnings of modern cosmology. The near future holds the promise 
of tightly constraining neutrino mass. 
However, the capacity of primary CMB alone to constrain ${\rm \sim Mpc}$ 
cosmology is rather limited; even CMB lensing by LSS, a sensitive 
probe of neutrino masses, takes place on considerably larger physical 
scales. In addition, standard forecasts of LE performance rely on the 
assumption that the primordial CMB signal is gaussian, and any 
nongaussianity present in the data should be attributed to lensing 
of the CMB. It should be emphasized that the results for neutrino masses 
derived from CMB and LE reported here, and in other works, 
always make this simplifying assumption. However, non-gaussianity can also 
be induced by astrophysical sources. Primordial nongaussianity is yet 
another source of confusion. On the relevant angular scales secondary CMB 
signals, such as the thermal and kinematic SZ effect, are known to be present. 
These are inherently nongaussian and can interfere with LE. If clustering of point 
sources is important on the relevant angular scales, then it is an additional 
nongaussian source that could be separated, in principle, from the lensed CMB 
signal using multifrequency observations, though this requires an accurate 
modeling of the emitting source, and would in any case be limited by the finite 
number of frequency channels.
 
Structure on Mpc scales probes the entire evolutionary 
history of matter perturbations down to these scales. This is especially 
relevant to neutrino physics via the effect of neutrino free streaming on 
these and larger scales. Indeed, Lyman-alpha ($Ly\alpha$) observations can provide 
additional diagnostic power to that from the CMB on these scales. However, since 
astrophysical conditions in the $Ly\alpha$ forest are not known very well, it is 
desirable to consider other probes of clustering on Mpc scales to complement 
$Ly\alpha$, as well as other traditional probes such as galaxy clustering, shear 
measurements, etc., as independent probes, and - minimally - to provide us with 
consistency checks. The possibility of using cluster catalogues to constrain 
neutrino masses has already been discussed by Wang et al. (2005). More generally, 
extraction of cosmological parameters, such as $\Omega_{b}$, $\Omega_{\Lambda}$, 
from cluster number counts - in conjugation with other cosmological probes - is widely 
discussed in the literature (e.g., Holder, Haiman \& Mohr 2001, Cunha, 
Huterer \& Frieman 2009, SSR).

Our results, summarized in Table 2, show that the projected uncertainties on neutrino 
masses lie in the $\sim 0.04-0.06$ eV range. 
This predicted relatively small mass uncertainty is competitive with that 
predicted in LE analyses.

The most important source of uncertainty in modeling cluster abundance and 
internal properties is the mass function. Current uncertainties in this 
basic function were explicitly included in our analysis. Continued extensive 
cosmological hydrodynamical simulations, e.g. Cunha \& Evrard (2009), are 
likely to result in a significantly more precise determination 
of this function across the cluster mass range. In contrast, uncertainties 
stemming from using simple models for the spatial profiles of the gas density 
and temperature are of secondary importance, simply because these are much-reduced 
when calculating integrated SZ measures and, more generally, 
have little effect on cluster detection. 
After all, the magnitude of the 
SZ-induced anisotropy is not determined by each of these quantities separately 
(in the non-relativistic limit that is valid for our purpose here), but rather by 
the integrated gas pressure along the line of sight. 

\section*{Aknowledgements}

We thank the referee for very constructive suggestions.
This work is supported in part by the Israel Science Foundation (grant
number 1362/08), by the European Research Council (grant number 203247), 
and by a grant from the James B. Ax Family Foundation.

\ \\
\ \\
\ \\

\begin{table}
\begin{tabular}{|c|c|c|c|c|}
\hline
Experiment & $f_{\rm sky}$ & $\nu [GHz]$ & $\theta_b [1']$ & $\Delta_T [\mu K]$\\
\hline
\hline
& &  30 & 33 &  4.4\\
& &  44 & 23 &  6.5\\
& &  70 & 14 &  9.8\\
& &  100 & 9.5 &  6.8\\
PLANCK&0.65& 143 & 7.1 & 6.0\\
& &  217 & 5.0 &  13.1\\
&   & 353 & 5.0 & 40.1\\
& &  545 & 5.0 &  401\\
&   & 856 & 5.0 & 18300\\
\hline
\hline
\end{tabular}
\caption{Sensitivity parameters for Planck: Only the 100, 143 and 353 GHz bands are 
used in the number counts estimates presented in this work.}
\end{table}

\begin{table}
\begin{tabular}{|c|c|c|c|c|c|c|}
\hline
Experiment&mass function&$\sigma_{M_{\nu}}[eV]$(prim.)&$\sigma_{M_{\nu}}[eV]$(LE)
&$\sigma_{M_{\nu}}[eV]$[prim.+N(z)]&$\sigma_{M_{\nu}}[eV]$[LE+N(z)]& $N_{clus}$\\
&uncertainty \%  &&&&&\\
\hline\hline
&0&&&0.06&0.06&\\
PLANCK&3&0.43&0.15&0.07&0.06&6040\\
&5&&&0.08&0.07&\\
&10&&&0.12&0.09&\\
\hline\hline
&0&&&0.04&0.03&\\
CVL&3&0.29&0.05&0.06&0.04&13860\\
&5&&&0.07&0.04&\\
&10&&&0.11&0.05&\\
\hline\hline
\end{tabular}
\caption{Statistical uncertainty on total neutrino mass from cluster number counts 
obtained from the PLANCK and CVL SZ surveys. Shown are the expected $1\sigma$ 
neutrino mass uncertainties. 
We show the results from CMB alone, LE,
CMB+cluster number counts N(z), and LE+N(z). 
The total cluster number expected to qualify into 
the sample is shown on the right column.}
\end{table}

\end{document}